\documentclass[%
 reprint,
superscriptaddress,
showpacs,preprintnumbers,
amsmath,amssymb,
aps,
pra,
]{revtex4-1}

\usepackage{graphicx}
\usepackage{dcolumn}
\usepackage{bm}
\usepackage{verbatim}

\begin{document}

\preprint{APS/123-QED}

\title{Discrete-time quantum walk approach to  high-dimensional quantum state transfer and quantum routing}

\author{Heng-Ji Li}
\affiliation{%
Information Security Center, State Key Laboratory Networking and Switching Technology, Beijing University of Posts Telecommunications, Beijing 100876, China}
\affiliation{%
School of Computer Science, Beijing University of Posts Telecommunications, Beijing 100876, China}


\author{Jian Li}
 \email{Email addresses:lijian@bupt.edu.cn}
\affiliation{%
School of Computer Science, Beijing University of Posts Telecommunications, Beijing 100876, China}

\author{Xiu-Bo Chen}
\affiliation{%
Information Security Center, State Key Laboratory Networking and Switching Technology, Beijing University of Posts Telecommunications, Beijing 100876, China}
\affiliation{%
GuiZhou University, Guizhou Provincial Key Laboratory of Public Big Data, Guizhou Guiyang, 550025, China}%
\date{\today}

\begin{abstract}
High-dimensional  quantum systems   can offer  extended possibilities and multiple advantages while developing advanced quantum technologies. In this paper, we propose a class of quantum-walk architecture networks that admit the efficient routing of high-dimensional quantum states.  Perfect state transfer of an arbitrary unknown qudit state can be achieved between two arbitrary nodes  via a one-dimensional lackadaisical discrete-time quantum walk. In addition,  this method can be generalized  to the high-dimensional lattices, where it allows distillable entanglement to be shared  between arbitrary input and output ports. Implementation of our scheme is more feasible through exploiting the coin degrees of freedom   and the settings of the coin flipping operators are simple.  These results provide  a direct application  in a high-dimensional computational architecture to process much more information.
\end{abstract}

\pacs{Valid PACS appear here}
\maketitle


\section{Introduction}
\label{intro}
High-dimensional  quantum systems (qudits) has emerging as an alternative to two-dimensional quantum systems (qubits), because they can offer  extended possibilities and multiple advantages while developing advanced quantum technologies.  For computation, implementing qudits brings about efficient distillation of resource states \cite{campbell2012magic} and simplified gates \cite{lanyon2009simplifying}. For quantum communication, qudits can lead to both higher information capacity \cite{erhard2018twisted} and increased noise resilience \cite{bechmann2000quantum}. The extensions of various protocols from qubits to qudits have been proposed, such as  universal quantum computation \cite{wang2020qudits}, quantum cryptography  \cite{wang2019characterizing} and even implemented in experiment \cite{hu2020experimental}. Additionally, the application of qudits will enhance and deepen our understanding of quantum computation and communication.

Quantum state transfer \cite{bose2003quantum,christandl2004perfect,
christandl2005perfect,wojcik2007multiuser,
chudzicki2010parallel} between two selected nodes  is a crucial task for quantum science technologies. The ability to perfectly transfer an arbitrary quantum state between different parts in the interior of quantum computers \cite{erhard2019characterizing} is essential and it is interesting from the perspective of distributed quantum computation \cite{cohen2018deterministic} while combining local quantum processing with quantum state transfer.  In addition, quantum state transfer  can be used to realize the distribution of arbitrary unknown multi-particle entanglement state in quantum network, which opens the possibility for a variety
of novel applications ranging from teleportation \cite{sun2016quantum}, error correction \cite{mazurek2020quantum}, and purification \cite{hu2021long}.

In this paper we consider two related fields of research: lackadaisical quantum walks (LQWs) on one-dimensional or higher-dimensional lattices and quantum state
transfer. In particular we are interested in
the discrete-time quantum walk realization of a high-dimensional quantum state
transfer sin a quantum-walk architecture.
Lackadaisical quantum walks(LQWs) was proposed by Wong \cite{wang2017one}, as the generalization of the original quantum walks(QWs), in which each vertex in a standard quantum walks is attached to $l$ self-loops. QWs, as the quantum mechanical analogs of classical random walks,  were first introduced by Aharonov et al. in $1993$ \cite{Aharonov1993Quantum}. Due to the  quantum interference effects, they  are computationally more efficient than their classical counterparts and offer an alternative approach to implementing better quantum algorithms, such as database search  \cite{Shenvi2003Quantum,Ambainis2005Coins,tulsi2008faster}, element distinctness \cite{ambainis2007quantum}, graph isomorphism \cite{gamble2010two} and so on. Later it was shown that they are capable of universal  quantum computation \cite{Childs2009Universal,Lovett2010Universal,childs2013universal}.

Some work \cite{kurzynski2011discrete,zhan2014perfect,
yalccinkaya2015qubit} on quantum state transfer has been investigated showing the promising application of QWs.  One approach to the problem is that one  performs the local coin operators at each individual node  with the full control of the walk-coin system. It is essentially the discrete-time variant of the engineered coupling protocol \cite{christandl2004perfect} in spin chains. They have shown that an arbitrary qubit  can be transferred with unit fidelity over arbitrary distance on the line \cite{zhan2014perfect}, cycle \cite{kurzynski2011discrete,yalccinkaya2015qubit} and square lattice \cite{zhan2014perfect}. In our work, we focus on the high-dimensional state transfer.

In this manuscript,  we propose a class of quantum-walk architecture networks that admit an efficient quantum routing,  where an arbitrary unknown qudit can be perfectly transferred   between
arbitrary input and output ports.
Consequently, sharing entangled qudits  to multiple arbitrary nodes can be achieved by extending  the method.  For achieving it,   we need to perform inhomogeneous coin flipping operators at every step, which are high-dimensional quantum gate: the identity operator, the generalized Pauli gate and  swap gate. It is shown that the time scaling between arbitrary sites is linear to the distance to be covered. By introducing the coin state, transferring the qudit is more feasible and easier to extend to multiqudit entanglement transfer.

The paper is structured as follows.  In Sec.\ref{preliminaries} the  preliminaries are provided about the knowledge of one-dimensional LQW on the line and three high-dimensional coin operators that are crucial for realizing our scheme.  In Sec.\ref{Qudits}, the scheme of transferring an arbitrary qutrit is presented and then it can be generalized  to the transfer of an arbitrary qudit.  In Sec.\ref{Nqudits}, We extend the scheme from one-dimensional quantum routing to the $m$-dimensional
case, routing entangled qudits to $m$ arbitrary
positions. Finally, Sec.\ref{Conclusion} contains our conclusions.

\section{preliminaries}
\label{preliminaries}
\subsection{One-dimensional LQWs on the line}
Lackadaisical quantum walks(LQWs) on the line was first introduced  in 2015 \cite{wang2017one}, as quantum analogue of lazy random walks where each vertex is attached to $\lambda$ self-loops. It is defined as a quantum system with two Hilbert spaces, the coin space $\mathcal{H}_c$
spanned by $d\,(d=\lambda+2)$ basis states $\{\left|0\right\rangle,\left|1\right\rangle,\cdots,\left|d-1\right\rangle\}$, and the position Hilbert space $\mathcal{H}_p$ spanned by $\{\left|x\right\rangle |x \in \mathbb{Z} \}$. The whole system is
in the space $\mathcal{H}=\mathcal{H}_p\otimes\mathcal{H}_c$.

One-step time evolution  of QWs is controlled by the unitary operator $\mathcal{U}=\mathcal{S}(\mathcal{I}\otimes \mathcal{C})$,
where $\mathcal{C} \in \mathcal{S}\mathcal{U}(d)$ is the coin flipping operator and $\mathcal{S}$ is the conditional shift operator described  the following unitary operator
\begin{equation}\label{shiftline}
\mathcal{S}\!=\!\sum_x \left|x\!-\!1,0\right\rangle\left\langle x,0\right|+\Big(
\sum_{i=1}^{\lambda}\left|x,i\right\rangle\left\langle x,i\right|\Big)
+\left|x\!+\!1,\lambda+1\right\rangle\left\langle x,\lambda+1\right|.
\end{equation}
where the index $x$ runs over $\mathbb{Z}$.  One finds clearly that the coin state $|0\rangle$ and $|d-1\rangle$ correspond to
the left and right and that the coin state $|i\rangle$ corresponds to the
neutral state for the motion and denote the direction of the walk as $\{l,s,r\}$.  An illustrative example (taking $\lambda=1$) is given in Figure\ref{lazzy}. If the particle and coin start in state
 $\left|\psi_0\right\rangle$, the state of the system after $t$ steps
 of the walk is  $\left|\psi_{t}\right\rangle=\mathcal{U}^{t}\left|\psi_{0}\right\rangle$.

\begin{figure}\label{lazzy}
\centering
\includegraphics[width=21pc]{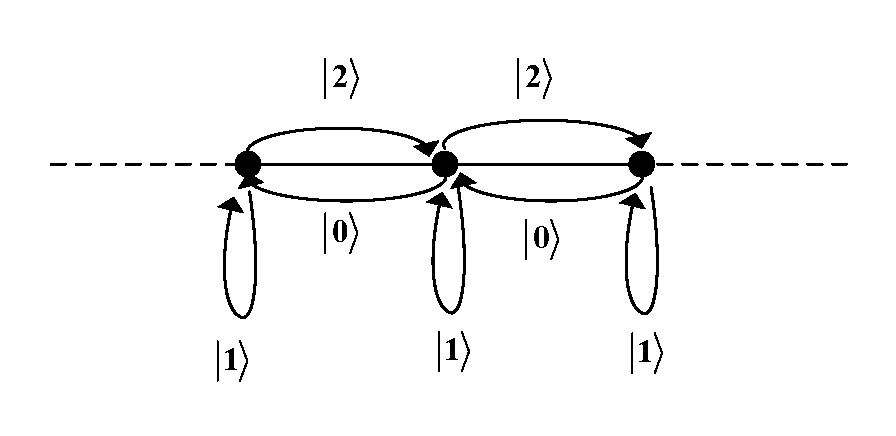}
\centering
\caption{The one-dimensional LQWs with one self-loop on the line. The quantum
state $\left|0\right\rangle$, $\left|1\right\rangle$ and  $\left|2\right\rangle$ determines  that the walker move left, stay put and move right, respectively.}\label{lazzy}
\label{figure0}\vspace*{-6pt}
\end{figure}

\subsection{High-dimensional quantum gate}
Before we show our scheme, we  introduce three kinds of  high-dimensional quantum gate performed on the coin state, the identity operator $\mathcal{I}$, the generalized Pauli gate $\mathcal{X}$ which is also called increment gate with $\mathcal{X}=\left|(k+1) \,\text{mod} \,d\right\rangle \left \langle k\right|$ and  swap gate $\mathcal{X}_{i\leftrightarrow j}$ \cite{di2013synthesis}, which act on the two-dimensional subspace $\mathcal{H}_{i,j}$ of $d$-dimensional Hilbert space with $\mathcal{X}_{i \leftrightarrow j}=|i\rangle \langle j|+|j\rangle \langle i|+\sum_{k\neq i,j}+|k\rangle \langle k|$. We call the operators $\mathcal{X}$ and $\mathcal{X}_{i \leftrightarrow j}$ as the \emph{special} coin operators.  For the identity coin operator $\mathcal{I}$, the coin states remain unchanged and therefore the direction of the walker remains the same as that for the previous step. For the special coin operators, it will change the coin state with the following forms
\begin{equation}
\left\{
\begin{aligned}
&|0\rangle \rightarrow |i\rangle(l\rightarrow s),|0\rangle \rightarrow |d-1\rangle(s\rightarrow r), \\
&|i\rangle \rightarrow |0\rangle(s\rightarrow l), |i\rangle \rightarrow |j\rangle(s\rightarrow s),|i\rangle \rightarrow |d-1\rangle(s\rightarrow r),\\
&|d-1\rangle \rightarrow |0\rangle(r\rightarrow s),|d-1\rangle \rightarrow |0\rangle(r\rightarrow l).
\end{aligned}
\right.
\end{equation}

\section{Transferring an arbitrary qudit via one-dimensional LQW}\label{Qudits}
\subsection{Perfect state transfer of an unknown qutrit}
To begin with, we introduce a basic scheme to transfer the three-dimensional coin state also called qutrit $|\Phi\rangle_{0}=\alpha \left|0\right \rangle+\beta \left|1\right\rangle+\gamma \left|2\right\rangle$, where $\alpha$, $\beta$, and $\gamma$ are complex numbers that fulfill $|\alpha|^{2}+|\beta|^{2}+|\gamma|^{2}=1$. For accomplishing it, take $\lambda=1$ in the shift operator (\ref{shiftline}).

Our goal is to transfer the coin state $|\Phi\rangle_{0}$ to a certain position $p$ from the original position $0$ after $n$-step QWs
\begin{equation}
|\Psi\rangle_{0}=|0\rangle|\Phi\rangle_{0}\stackrel{n\, \text{steps}}{\longrightarrow}|\Psi\rangle_{n}=|p\rangle|\Phi\rangle_{0},
\end{equation}
where $\left|\Psi\right\rangle_{i}$ denotes quantum state after the $i$ steps. Consequently,  it can be derived that
\begin{equation}\label{initial3n1}
\begin{split}
\left|\Psi\right\rangle_{n-1}=\sum_{j=-1,0,1} a_{n-1,p+j}\left|p+j\right\rangle\left|\phi\right\rangle_{n-1,p+j},
\end{split}
\end{equation}
where $\left|a\right\rangle_{n,p}$ and $\left|\phi\right\rangle_{n,p}$, respectively, are the complex amplitude and the coin state corresponding to the walker in position $p$ after the $n$-th step. The position state in $\left|\Psi\right\rangle_{n-1}$ is $p-1$, $p-1$ or $p$,  due to the fact that the
states in other positions cannot walk to the position in $p$ in one step.

Next, by using  the iteration relation
\begin{equation}\label{iteration}
\begin{split}
\left|\Psi\right\rangle_{n}=\mathcal{S}\sum_{x}(\mathcal{I}\otimes\mathcal{C}_{n,x})\left|\Psi\right\rangle_{n-1,x},
\end{split}
\end{equation}
where $\mathcal{C}_{n,x}$ stands for the coin flipping operator performed on the position $x$ of the $n$-th step, and substituting (\ref{shiftline}) and (\ref{initial3n1}) into (\ref{iteration}), it can be concluded that
\begin{equation}
\left\{
\begin{aligned}
&a_{n-1,p+1}=\alpha,\, \mathcal{C}_{n,p+1} \left|\phi\right\rangle_{n-1,p+1}=\left|0\right\rangle,\\
&a_{n-1,p}=\beta, \,\mathcal{C}_{n,p} \left|\phi\right\rangle_{n-1,p}=\left|1\right\rangle, \\
&a_{n-1,p-1}=\gamma,\, \mathcal{C}_{n,p-1} \left|\phi\right\rangle_{n-1,p-1}=\left|2\right\rangle.
\end{aligned}
\right.
\end{equation}

By analysis,  we will take $\mathcal{C}_{1,0}=\mathcal{X}_{0\leftrightarrow2}$ to swap the information flow $\alpha$ and $\gamma$, and therefore after the first step the whole walk-coin system will be
\begin{equation}
\left|\Psi\right\rangle_{1}=\gamma\left|-1\right \rangle \left|0\right \rangle+\beta \left|0\right \rangle\left|1\right\rangle+\alpha \left|1\right \rangle\left|2\right\rangle
\end{equation}

Thus, we need to determine the coin unitary operators from second to $(n-1)$-th step
\begin{equation}
|\Psi\rangle_{1}\stackrel{\mathcal{C}_{2,x}}{\longrightarrow}|\Psi\rangle_{2}\dashrightarrow
|\Psi\rangle_{n-2}\stackrel{\mathcal{C}_{n\!-\!1,x}}{\longrightarrow}|\Psi\rangle_{n-1}
\end{equation}
which can achieve the goal, propagating to position $x-1$, $x$, or $x+1$ from position $-1$, $0$, or $1$ after the $(n-1)$-th step and correspondingly, we need to realize the transmission of the  information flow of $\alpha$, $\beta$ and $\gamma$ as follows
\begin{equation}
\left\{
\begin{aligned}
&\alpha \left|1\right \rangle\left|2\right\rangle\rightarrow\alpha \left|p+1\right \rangle\left|\phi\right\rangle_{n-1,p+1},\\
&\beta \left|0\right \rangle\left|1\right\rangle\rightarrow\beta \left|p\right\rangle  \left|\phi\right\rangle_{n-1,p},\\
&\gamma\left|-1\right \rangle \left|0\right \rangle\rightarrow \gamma\left|p-1\right \rangle  \left|\phi\right\rangle_{n-1,p-1}.
\end{aligned}
\right.
\end{equation}

For solving it, the set of  the coin unitary operator used will be  $\{\mathcal{I}$,  $\mathcal{X}$, $\mathcal{X}^{2}\}$ and   by using the model of the  classical lazy random walks, it can be derived
\begin{equation}\label{lsr}
\left\{
\begin{aligned}
&n_{l}+n_{s}+n_{r}=n-2,\\
&n_{r}-n_{l}=p,
\end{aligned}
\right.
\end{equation}
where $n_{l}$, $n_{s}$ and $n_{r}$ ,respectively, are the number of moving left, staying put, and moving right which will give
\begin{equation}
  n-2\geq |p|.
\end{equation}
which restricts the range of the position that the coin state can be perfectly transferred.
Furthermore, it will yield
\begin{equation}
\left\{
\begin{aligned}
&n_{l}=\frac{n-2-n_{s}-p}{2},\\
&n_{r}=\frac{n-2-n_{s}+p}{2},\\
&n_{s}=n-2-|p|-2n_{\Delta}.
\end{aligned}
\right.
\end{equation}
where $n_{\Delta}=\text{min}\{n_{r},n_{l}\}$. Hence the number of such walks that satisfy the equation (\ref{lsr}) will be
\begin{equation}
N(n,p)=\sum_{n_{s}=0}^{n-2-|p|}\binom{n-2}{n_{s}}\binom{n-2-n_{s}}{(n-2-n_{s}+p)/2},
\end{equation}
which is depending on the step $n$ and the target position $p$. Correspondingly there will be $N(n,p)$ kinds of coin unitary operators and denote the solution space as $\Omega=\{\tau_{1},\tau_{2}\cdots,\tau_{N(n,p)}\}$.

Denote $C(\tau_{i})$ as the sum of  the numbers of the special coin operators $\{\mathcal{X},\mathcal{X}^2\}$ for the case $\tau_{i}$. The target is finding out $\tau_{i}^{*}$ that satisfies the condition
 \begin{equation}
 C(\tau_{i}^{*})=\text{min}\,C(\tau_{i}), {\forall} p_{i} \in \Omega.
\end{equation}

It can be easily obtained that the solution should have the following form
$$
\underbrace{d_{1}d_{1}\cdots d_{1}}_{n_{d_{1}}}\underbrace{d_{2}d_{2}\cdots d_{2}}_{n_{d_{2}}}\underbrace{d_{3}d_{3}\cdots d_{3}}_{n_{d_{3}}}
$$
where $d_{1},d_{2},d_{3} \in \{l,s,r\}$ with $d_{1}\neq d_{2} \neq d_{3}$.

(i)The walking of the information flow $\alpha$ from $1$ to $p+1$, it can be derived that
$$
\underbrace{rr\cdots r}_{n_{r}}\underbrace{ss\cdots s}_{n_{s}}\underbrace{ll\cdots l}_{n_{l}}
$$
Due to the constraint $n+p=2(n_{r}+1+\frac{n_{s}}{2})$, while $n_{s}$ is even or odd, the parity of $n$ and $p$ will be the same or different. In order to transfer the state  to  position $p$ after an arbitrary $n$-step QW, both two cases where $n_{s}$ is even or odd need  to be considered.

$\textbf{Case 1}$: While $n_{s}$ is even, the  solution is that $\mathcal{X}$ needs to be performed to make $\left|2\right\rangle\rightarrow \left|0\right\rangle$  at the location $\frac{n+p}{2}$ in the $\frac{n+p+2}{2}$ step with $n_{s}=0$. It is because that if $n_{s}\neq0$, $\mathcal{X}^2$ will be  performed twice for making $\left|2\right\rangle\rightarrow \left|1\right\rangle$ and $\left|1\right\rangle\rightarrow \left|0\right\rangle$.

$\textbf{Case 2}$: While $n_{s}$ is odd, the solution is $\mathcal{X}^2$ needs to be performed respectively for achieving $\left|2\right\rangle\rightarrow \left|1\right\rangle$  and $\left|1\right\rangle\rightarrow \left|0\right\rangle$ at the position $\frac{n+p-n_{s}}{2}$ in the $\frac{n+p+2-n_{s}}{2}$ and $\frac{n+p+2+n_{s}}{2}$  steps.  Without loss of generality, take $n_{s}=n-2-|p|$ to achieve the transfer.

(ii)For the case that the walking of the information flow $\gamma$ from $-1$ to $p-1$, similarly it can be deduced that
$$
\underbrace{ll\cdots l}_{n_{l}}\underbrace{ss\cdots s}_{n_{s}}\underbrace{rr\cdots r}_{n_{r}}
$$
and then we will show the result directly.

$\textbf{Case 1}$: While $n_{s}$ is even,  the  solution is that $\mathcal{X}^2$ is needed  to render $\left|0\right\rangle\rightarrow \left|2\right\rangle$  at the location $\frac{p-n}{2}$ in the $\frac{n-p+2}{2}$ step with $n_{s}=0$.

$\textbf{Case 2}$: While $n_{s}$ is odd,  the solution is that $\mathcal{X}$ is used  to achieve $\left|0\right\rangle\rightarrow \left|1\right\rangle$ and  $\left|1\right\rangle\rightarrow \left|2\right\rangle$ at the location $\frac{p+n_{s}-n}{2}$ in the $\frac{n-p+2-n_{s}}{2}$  and $\frac{n-p+2+n_{s}}{2}$ step. And without loss of generality, take $n_{s}=n-2-|p|$ to achieve the transfer.

(iii)For the case that the walking of the information flow $\beta$ from $0$ to $p$, the solution will be
$$
\underbrace{ss\cdots s}_{j}\underbrace{ww\cdots w}_{|p|}\underbrace{ss\cdots s}_{n_{s}-j}
$$
where $n_{s}=n-2-|p|$, $0\leq j \leq n_{s}$ and
$w=\left\{
\begin{aligned}
&l, p<0;\\
&r, p>0
\end{aligned}
\right.
$. Therefore, at position $0$ and $p$ in the $j+2$  and $j+2+|p|$ step, (i) for $p>0$,
$\mathcal{X}$ and $\mathcal{X}^{2}$
 are respectively performed to make $\left|1\right\rangle\rightarrow \left|2\right\rangle$ and  $\left|2\right\rangle\rightarrow \left|1\right\rangle$; (ii) for $p<0$,
$\mathcal{X}^{2}$ and $\mathcal{X}$
 are respectively performed to make $\left|1\right\rangle\rightarrow \left|0\right\rangle$ and  $\left|0\right\rangle\rightarrow \left|1\right\rangle$.
Without loss of generality, take $j=0$ to achieve the transmission task.

To sum up, the transmission of the information flow $\alpha$, $\beta$ and $\gamma$ have been achieved. The setting of the special coin flipping operators depends on the target position and the step
numbers and they are as follows with leaving
the others equal to $\mathcal{I}$. There are 5 or 7 special coin operators while the parity of $n$ and $p$ are  the same or different. Specially, the common coin flipping operators for achieving the transfer of the information flow $\beta$ are shown below.
\begin{equation}
\mathcal{X}_{0\leftrightarrow2}:(1,0),
\mathcal{X}:
(|a_{2}|\!+\!2,a_{2}),
\mathcal{X}^{2}:(|a_{1}|\!+\!2,a_{1});\\
\end{equation}
As for achieving the transfer of the information flow $\alpha$ and $\gamma$, there are two different sets of the coin operators.

(i)While the parity of $n$ and $p$ are the same:
\begin{equation}
\mathcal{X}:\,(b^{+}+1,b^{+}),\,\,\mathcal{X}^{2}:(b^{-}+1,-b^{-});
\end{equation}

\begin{figure}
\centering
\includegraphics[width=21pc]{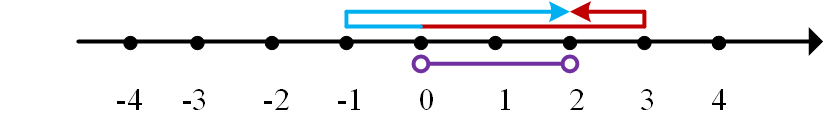}
\centering
\caption{Perfect state transfer of the coin state $\alpha \left|0\right \rangle+\beta \left|1\right\rangle+\gamma \left|2\right\rangle$ from position 0 to 2 after four-step discrete-time walks. The  red, purple and blue arrows indicate the directions of the information flow of $\alpha$, $\beta$ and $\gamma$, respectively.}
\label{Figure1a}\vspace*{-6pt}
\end{figure}

\begin{figure}
\centering
\includegraphics[width=21pc]{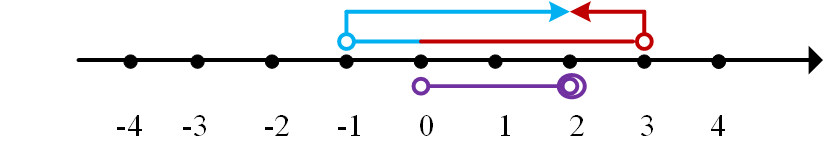}
\centering
\caption{Perfect state transfer of the coin state $\alpha \left|0\right \rangle+\beta \left|1\right\rangle+\gamma \left|2\right\rangle$ from position 0 to 2 after five-step discrete-time walks. The  red, purple and blue arrows indicate the directions of the information flow of $\alpha$, $\beta$ and $\gamma$, respectively.}
\label{Figure1b}\vspace*{-6pt}
\end{figure}

(ii) While the parity of $n$ and $x$ are different:
\begin{equation}
\left\{
\begin{aligned}
&\mathcal{X}^{2}:
(|a_{1}|\!+\!2,\!a_{1}\!+\!1),(n-|a_{2}|,a_{1}\!+\!1);\\
&\mathcal{X}:\,\,\,(|a_{2}|\!+\!2,a_{2}\!-\!1),\,(n-|a_{1}|,a_{2}\!-\!1).
\end{aligned}
\right.
\end{equation}
where $(,)$  is the two-dimensional array of the step and position, $a_{i}=p\delta_{i}(p)(i=1,2)$, $b^{\pm}=\frac{n\pm p}{2}$ $\delta_{1}(p)=\left\{
\begin{aligned}
&1,p>0;\\
&0,p<0
\end{aligned}
\right.
$
and $\delta_{2}(p)=\left\{
\begin{aligned}
&0, p>0;\\
&1, p<0.
\end{aligned}
\right.
$ Two examples of transferring three-dimensional quantum state from position 0 to
position 2 via a four and five-step discrete-time QW are shown in Figure \ref{Figure1a} and \ref{Figure1b}.

By taking $\lambda=0$, it can be  deduced that the two-dimensional state (qubit) $|\Phi\rangle_{0}=\alpha \left|0\right \rangle+\gamma \left|1\right\rangle$ can  be transferred to target position $p$ after $n$-step walks with the  coin unitary operator $\sigma_{x}$ at $(1,0)\,,(b^{-}+1,-b^{-})\,,(b^{+}+1,b^{+})$. It should be noted that for even
(odd) step numbers the coin state can be transferred only to the even (odd) positions because of the lacking of ``staying put".

\subsection{Perfect state transfer of an unknown qudit}
We now consider the transfer of arbitrary $d$-dimensional quantum state
\begin{equation}
|\Phi\rangle_{0}=\alpha \left|0\right \rangle+\sum_{i=1}^{d-2}\beta_{i} \left|i\right\rangle+\gamma \left|d-1\right \rangle,
\end{equation}
where  $\alpha$, $\beta_{i}$
and $\gamma$ are complex numbers, and $|\alpha|^{2}+\sum_{i}^{d-2}|\beta_{i}|^{2}+|\gamma|^{2}=1$.
We will take  $\mathcal{C}_{1,0}=\mathcal{X}_{0\leftrightarrow d-\!1}$,
and it will yield
\begin{subequations}
\begin{align}
\left|\Psi\right\rangle_{1}=&\gamma\left|-1\right \rangle \left|0\right \rangle+\sum_{i=1}^{d-2}\beta_{i}
\left|0\right \rangle\left|i\right\rangle+\alpha \left|1\right \rangle\left|d-1\right\rangle,\\
\left|\Psi\right\rangle_{n-1}=& \alpha\left|p+1\right\rangle\left|\phi\right\rangle_{n-1,p+1}
+\gamma\left|p-1\right\rangle\left|\phi\right\rangle_{n-1,p-1}+\notag\\
&\sum_{i=1}^{d-2}\beta_{i}\left|p\right\rangle|\phi^{(i)}\rangle_{n-1,p}.
\end{align}
\end{subequations}

Then take the following coin unitary operators in  the $n$-th step
\begin{equation}
\left\{
\begin{aligned}
&\mathcal{C}_{n,p+1} \left|\phi\right\rangle_{n-1,p+1}=\left|0\right\rangle,\\
&\mathcal{C}_{n,p}\sum_{i=1}^{d-2}\beta_{i}|\phi^{(i)}\rangle_{n-1,p}=\sum_{i=1}^{d-2}\beta_{i}|i\rangle, \\
&\mathcal{C}_{n,p-1} \left|\phi\right\rangle_{n-1,p-1}=\left|d\!-\!1\right\rangle.
\end{aligned}
\right.
\end{equation}
and thereby it will deduce that
\begin{equation}
\left|\Psi\right\rangle_{n-1}\stackrel{\mathcal{C}_{n}} {\longrightarrow} \left|\Psi\right\rangle_{n}=\left|p\right \rangle|\phi\rangle_{0}.
\end{equation}

Therefore, our goal is  to realize the transmission of the  information flow of $\alpha$, $\beta_{i}$ and $\gamma$ as follows
\begin{equation}
\left\{
\begin{aligned}
&\alpha \left|1\right \rangle\left|d-1\right\rangle\rightarrow\alpha \left|p+1\right\rangle \left|\phi\right\rangle_{n-1,p+1},\\
&\beta_{i} \left|0\right \rangle\left|i\right\rangle\rightarrow\beta_{i}|p\rangle|\phi^{(i)}\rangle_{n-1,p},\,1\leq i\leq d-2,\\
&\gamma\left|-1\right \rangle \left|0\right \rangle\rightarrow \gamma\left|p-1\right \rangle \left|\phi\right\rangle_{n-1,p-1}.
\end{aligned}
\right.
\end{equation}

Considering the walking of the information
flow $\alpha$ and $\gamma$, the corresponding solution is the same as the transfer of qutrit. Consequently,  our work is gain access to achieving the  transmission of the information $\beta_{i}$ from $0$ to $x$. The main idea is to perform the proper  coin unitary operator at the original position to break the coherence. The detailed process  is shown below.

(i)Keep performing  the coin unitary operator $\mathcal{X}^{k(p)}$ at position $0$ from second to $(d\!-\!1)$-th step and in the $j$-th step $(2\leq j \leq d-1)$ it will make $\beta_{f(j,p)}|f(j,p)\rangle$ become  $\beta_{f(j,p)}|d(p)\rangle$  with $k(p)=\delta_{1}(p)+(d-1)\delta_{2}(p)$, $f(j,p)=(d-j)\delta_{1}(p)+(j-1)\delta_{2}(p)$ and $d(p)=(d-1)\delta_{1}(p)$.

(ii)While $\beta_{f(j,p)}|d(p)\rangle$ arriving at position $p$ one by one, $\mathcal{X}_{d(p)\leftrightarrow f(j,p)}$ is performed one by one
and it will make $\beta_{f(j,p)}|d(p)\rangle$ become $\beta_{f(j,p)}|f(j,p)\rangle$.  As a result, we achieve the transfer  of  to position $p$ of all the information flow of $\beta_{i}$, and it needs to meet the condition $n\geq |p|+d-1$.

Next, we will show the coin flipping operators.  The common special coin operators which can achieve the transfer of the information flow $\beta_{i}$ are shown below.
\begin{equation}\label{c1qudit}
\mathcal{X}_{0\leftrightarrow d-1}:(1,0);
\mathcal{X}^{k(p)}\!:(j,0);\mathcal{X}_{d(p)\leftrightarrow f(j,p)}:(|p|+j,p);\\
\end{equation}
And two cases are  considered below, in order to achieve the transfer of the information flow $\alpha$ and $\gamma$.

(i)While the parity of $n$ and $p$ are the same:
\begin{equation}\label{c2qudit}
\mathcal{X}:\,(b^{+}+1,b^{+}),\,\,\mathcal{X}^{d-1}:(b^{-}+1,-b^{-});
\end{equation}

(ii) While the parity of $n$ and $x$ are different:
\begin{equation}\label{c3qudit}
\left\{
\begin{aligned}
&\mathcal{X}^{2}:
(|a_{1}|\!+\!2,\!a_{1}\!+\!1),\mathcal{X}^{d-1}:(n-|a_{2}|,a_{1}\!+\!1);\\
&\mathcal{X}:
(|a_{2}|\!+\!2,a_{2}\!-\!1),\mathcal{X}^{d-2}:\,(n-|a_{1}|,a_{2}\!-\!1).
\end{aligned}
\right.
\end{equation}

\begin{figure}
\centering
\includegraphics[width=21pc]{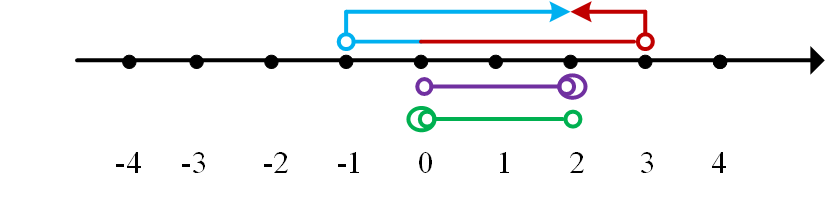}
\centering
\caption{Perfect state transfer of the coin state $\alpha \left|0\right \rangle+\beta_{1} \left|1\right\rangle+\beta_{2} \left|2\right\rangle+\gamma \left|3\right\rangle$ from position 0 to 2 after five-step discrete-time walks. The  red, green,  purple and blue arrows indicate the directions of the information flow of $\alpha$, $\beta_{1}$,  $\beta_{2}$ and $\gamma$, respectively.}
\label{Figure2}\vspace*{-6pt}
\end{figure}

In summary,  in our scheme $2d-1$ or $2d+1$ special coin  operators are needed depending on the parity of $n$ and $p$, meaning that the number of the special operators grows linearly as the dimensionality increases. An examples of transferring the four-dimensional quantum state from position 0 to
position $2$ via a five-step discrete-time QW is shown in Figure \ref{Figure2}.

\section{Routing multiqudit entangled state via high-dimensional QWs} \label{Nqudits}

For realizing the transferring  of an arbitrary unknown $m$-qudit entangled state, we now consider the $m$-dimensional discrete-time QWs, in which $m$ coins control the walkers to walk in different directions. Our goal is to achieve the transfer from the initial position $(0,\cdots,0)$ to target position $\textbf{\emph{p}}=(p_{1},\cdots,p_{N})$ after $n$-step walks. It also means the unknown multi-particle high-dimensional quantum state can be successfully
delivered to the users corresponding to the positions. While the unknown quantum state is entangled, a quantum information process based on entanglement distribution can be carried out successfully  after efficient routing. For the sake of simplicity, here we explain it using the example of a $2$-dimensional system.

Define the unitary operator of the $i$-th walk-coin system for achieving the transfer to position $p_{i}$ after $n$-step  as $\mathcal{U}_{i}(n,p_{i})$
 abbreviated as $\mathcal{U}_{i}$ and thus we can obtain
\begin{equation}
\begin{split}
\mathcal{U}_{1}|0\rangle|\Phi_{1}\rangle=|p_{1}\rangle|\Phi_{1}\rangle\\
\mathcal{U}_{2}|0\rangle|\Phi_{2}\rangle=|p_{2}\rangle|\Phi_{2}\rangle
\end{split}
\end{equation}
which will yield
\begin{equation}
(\mathcal{U}_{1}\otimes\mathcal{U}_{2})(|0,0\rangle|\Phi_{1}\rangle|\Phi_{2}\rangle)
=|p_{1},p_{2}\rangle|\Phi_{1}\rangle|\Phi_{2}\rangle
\end{equation}
where $|\Phi_{1}\rangle|\Phi_{2}\rangle$ is a separate state with the form $\sum_{i}^{d-1}a_{i}|i\rangle\sum_{j}^{d-1}a_{j}|j\rangle$.   Because  $\mathcal{U}_{i}$ is independent on the amplitude  of  state transferred,  we can rewrite as
\begin{equation}
(\mathcal{U}_{1}\otimes\mathcal{U}_{2})(|0,0\rangle|\Phi_{s}\rangle)
=|p_{1},p_{2}\rangle|\Phi_{s}\rangle
\end{equation}
where $|\Phi_{s}\rangle=\sum_{i}^{d-1}\sum_{j}^{d-1}a_{ij}|ij\rangle$ which is an arbitrary two-particle quantum state. It means that an arbitrary  quantum state $|\Phi_{s}\rangle$ can be transferred to the position $(p_{1},p_{2})$, by performing the unitary operator $\mathcal{U}_{1}\otimes\mathcal{U}_{2}$. The coin flipping operator in position $(x_{1},x_{2})$ at the $k$-th step will be $\mathcal{C}^{(1)}_{k,x_{1}}\otimes\mathcal{C}^{(2)}_{k,x_{2}}$ defined by the formulas (\ref{c1qudit}), (\ref{c2qudit}) and (\ref{c3qudit}), which is a local operation and does not break the entanglement between the two walk-coin systems. Especially, while taking $a_{ij}=0(i\neq j)$, the state will turn to be $\sum_{i}^{d-1}a_{i}|ii\rangle$, which is more interesting and significant for many quantum information precesses.

\begin{figure}
\centering
\includegraphics[width=21pc]{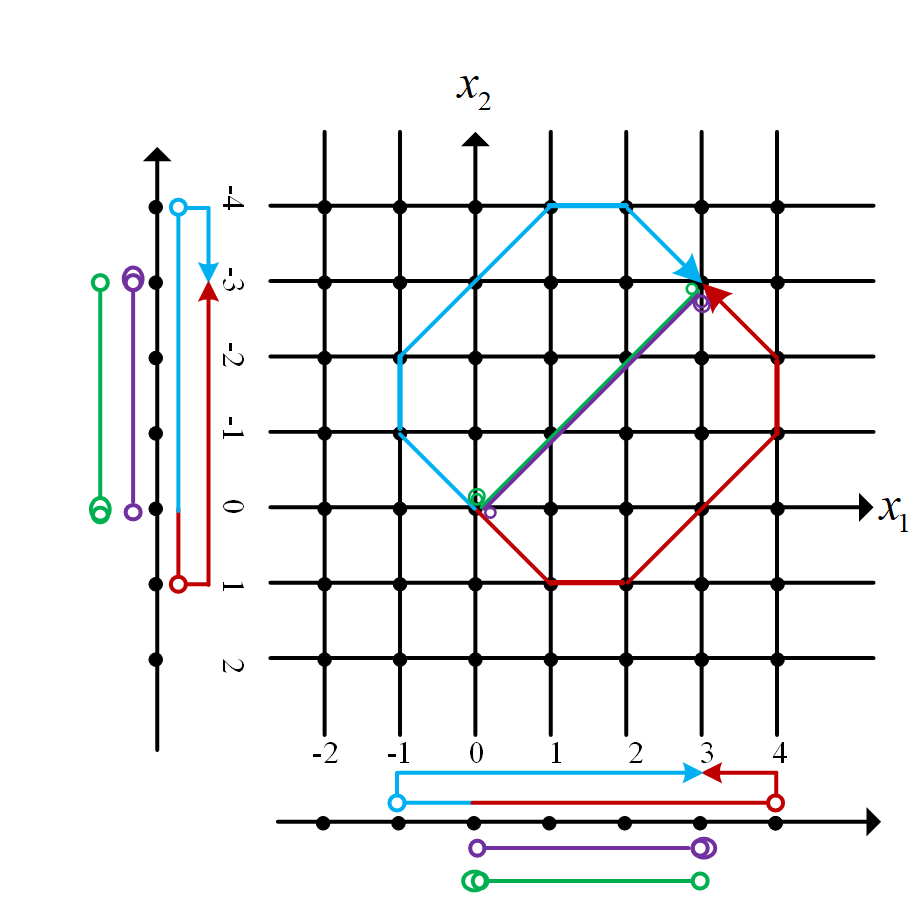}
\centering
\caption{Perfect state transfer of the coin state $\alpha|00\rangle+\beta_{1}|11\rangle
+\beta_{2}|22\rangle+\gamma|33\rangle$ from position $(0,0)$ to $(3,-3)$ after six-step walks in the two-dimensional case. The first
walker walks along the $x_{1}$ axis and the second walker walks along the
$x_{2}$ axis. The  red, green,  purple and blue arrows indicate the directions of the information flow of $\alpha$, $\beta_{1}$,  $\beta_{2}$ and $\gamma$, respectively.}
\label{Figure3}\vspace*{-6pt}
\end{figure}

For example, take $n=6$, $(p_{1},p_{2})=(3,-3)$ and the
coin operators in the first step will be
$\mathcal{C}^{(1)}_{1,0} \otimes
\mathcal{C}^{(2)}_{1,0}=\mathcal{X}_{0\leftrightarrow
d-1}\otimes\mathcal{X}_{0\leftrightarrow d-1}$.
An examples of transferring the four-dimensional quantum entanglement state $\alpha|00\rangle+\beta_{1}|11\rangle
+\beta_{2}|22\rangle+\gamma|33\rangle$ from position $(0,0)$ to
position $(3,-3)$ via a six-step discrete-time QW is shown in Figure \ref{Figure3}.

Furthermore, for multiple coins in the architecture of a $N$-dimensional discrete-time QW, the  efficient routing scheme  can be developed  based on the way we develop one dimensional system to two-dimensional system. Then it can be easily generalized to the  $m$-dimensional case, that is
\begin{equation}
\otimes_{i=1}^{m}\mathcal{U}_{i}(|0\rangle^{\otimes m}|\Phi_{s}^{m}\rangle)
=(\otimes_{i=1}^{m}|x_{i}\rangle)|\Phi_{s}^{m}\rangle
\end{equation}
where $|\Phi_{s}^{m}\rangle=\sum_{i_{1}}^{d-1}\cdots\sum_{i_{m}}^{d-1}a_{i_{1}\cdots i_{m}}|i_{1}\cdots i_{m}\rangle$ and it means that an arbitrary $m$-particle quantum state $|\Phi_{s}^{m}\rangle$ can be transferred to the arbitrary position $\textbf{\emph{p}}$, by performing the unitary operator $\otimes_{i=1}^{m}\mathcal{U}_{i}$.

For every walk-coin system, two kinds of special coin flipping operators are utilized that  depends on the target position and the step numbers. The time of routing
multiqudit quantum states between arbitrary sites is linear to the distance to be covered. Efficiently routing $m$ coins  needs $m(2d\pm1)$ special settings, which grows linearly with then number of the coins.

\section{Conclusion}
\label{Conclusion}
In this paper, we have demonstrated that an
arbitrary unknown qudit can  be transferred   with unit fidelity over arbitrary distances. This is a direct application to communicate between
two remote registers  in a computational architecture using high-dimensional
systems, which can construct a bigger Hilbert
space to process much more information than the two-dimensional
ones. One can perfectly transfer the unknown high-dimensional coin state in a one-dimensional quantum-walk architecture from the initial position to the target position.  $2d-1$ or $2d+1$ special coin  operators are needed depending on the parity of $n$ and $p$, while leaving
the others equal to $\mathcal{I}$.  Consequently, routing multiqudit entanglement  can be realized based on the state transfer on the regular network.  Efficiently routing $m$ coins  needs $m(2d\pm1)$ special settings, which grows linearly with then number of the coins.  The settings of the scheme are simple and independent of the number of target positions, which
makes our protocol feasible with the current experimental
technology. 
 
\bibliography{PSTReferences}

\end{document}